\documentclass[]{interact}

\usepackage{epstopdf}
\usepackage[caption=false]{subfig}

\usepackage[numbers,sort&compress]{natbib}
\bibpunct[, ]{[}{]}{,}{n}{,}{,}
\makeatletter
\def\NAT@def@citea{\def\@citea{\NAT@separator}}
\makeatother

\theoremstyle{plain}

\theoremstyle{definition}

\theoremstyle{remark}

\begin{document}


\title{Phase randomization in disordered layered systems}

\author{
\name{R.~S. Puzko\textsuperscript{a,b}\thanks{CONTACT R.~S. Puzko. Email: roman998@mail.ru} and A.~M. Merzlikin\textsuperscript{a,b,c}}
\affil{\textsuperscript{a}Institute for Theoretical and Applied Electrodynamics, Russian Academy of Sciences, 13 st. Izhorskaya, Moscow, 125412, Russia; \textsuperscript{b}Dukhov Automatics Research Institute (VNIIA), 22 st. Sushchevskaya, Moscow, 127055, Russia; \textsuperscript{c}Moscow Institute of Physics and Technology, 9 Institutskiy per., Dolgoprudny, Moscow Region, 141701, Russia}
}

\maketitle

\begin{abstract}
The propagation of light through a disordered layered system is studied. It is shown that distribution function of the transmission coefficient phase tends to stationary non-uniform distribution as the number of layers increases. The exponential convergence to the stationary distribution allows unambiguous definition of the phase randomization length scale.
\end{abstract}


\section{Introduction}

The wave propagation through disordered systems was actively studied since the middle of the 20th century~\cite{1abrahams2010,2sheng2006,3lifshits1982}. P.W. Anderson in his pioneering paper~\cite{4anderson1958} showed the absence of electron diffusion in the one-dimensional chain of random potentials. The paper~\cite{4anderson1958} started the massive research of particle (or wave) localization in disordered systems.

The study of wave localization in disordered systems was initially developed in the framework of solid state physics. The phenomenon was transferred to electrodynamics due to the similarity of the wave equations (see for example~\cite{5goto2009}). Notably, the study of interference phenomena (including Anderson localization) in electrodynamics has significant advantage due to the absence of interaction between photons.

The Anderson localization of light results in the exponential decrease of the transmission coefficient $t$ with the length $L$ of a disordered layered system~\cite{2sheng2006}
\begin{eqnarray}
t=\left|t\right|e^{i\varphi},
\label{eq:transmission}
\\
\left|t\right|\sim e^{-L/L_{loc}},
\label{eq:Lloc}
\end{eqnarray}
where $L_{loc}$ is the Anderson localization length and $\varphi$ is the transmission coefficient phase.

Due to the complexity of analytical approach the first theoretical models of Anderson localization were based on a priori assumptions. Now the numerical experiments allow direct simulation of the Anderson localization. As a result some a priori assumptions were strongly revised (including the scaling hypothesis~\cite{6luan2001}).

The phase randomization approximation (PRA)\footnote{As a rule, the phase of transmission or reflection coefficient is considered in various approaches.} is an assumption widely used in localization theory. In particular, the random T-matrix approach is based on PRA~\cite{7anderson1980,8lambert1982}. According to PRA, the distribution of phase $\varphi_{red}$ reduced to the interval $\left[0,2\pi\right]$ tends to the uniform distribution as the system length increases\footnote{The phase distribution calculated either for an ensemble of disordered systems at a fixed point or for different points of one extended system is considered. The distributions obtained in both methods converge according to the ergodic hypothesis.}. The convergence of the distribution occurs at a certain length $L_{ph}$ (phase randomization length), which is considered significantly smaller than the localization length.

A number of studies discussed the validity of PRA and its interpretation. In particular, studies~\cite{8lambert1982,9lambert1983} showed the violations of PRA in particular systems: the phase distribution converges to the non-uniform one in a number of cases. The non-uniform distribution of the phase relates with the correction to the localization length~\cite{8lambert1982,10izrailev1998}. The anomalies of Anderson localization scaling were found at certain energies (wavelengths) for particular disordered systems~\cite{11kappus1981,12derrida1984,13goldhirsch1994}. The violations of scaling at the gap edges for systems with weak disorder~\cite{10izrailev1998} relate to the non-uniform distribution of the phase. The study of scaling anomalies due to non-uniform distribution of the phase is given in~\cite{14deych1998,15liu2015,16tessieri2018,17torres2011,18kravtsov2011}. The numerical studies of the transmission coefficient~\cite{19lee2012} show that the phase distribution is a non-uniform one even in the limit of large size system. The phase of the reflection coefficient is studied as well. In particular, the non-uniform distribution of the reflection coefficient phase was demonstrated in particular systems~\cite{20kim1998,21titov2005}. Analytical studies~\cite{21titov2005} show the non-uniformity of phase distribution and the concurrent scaling violation in case of a system with a short-range correlated disorder. At the same time, the seemingly natural assumption of a uniform phase distribution is widely used~\cite{7anderson1980,22lazo2014,23king2017}.

Thus, there are several independent opinions in the literature on the phase randomization, which indicates the absence of a unified approach to this phenomenon. The question remains open whether the phase randomization exists at all. In other words, the PRA is accompanied by following unclear issues:

(a)	Which phase should be considered: the phase of transmission or reflection coefficient (or even their composition)?

(b)	Does the phase distribution converge to the uniform or non-uniform distribution?

(c)	Is there any length scale of the phase distribution convergence?

To answer the questions we conduct the numerical study of the phase randomization, verifying the convergence of the phase distribution to the non-uniform one. The revealed properties of this convergence allows for unambiguous definition of the phase randomization length $L_{ph}$. Moreover, the study shows the different frequency dispersions of phase randomization length and the localization length.

\section{Phase randomization}

Let us now consider a layered disordered system. The phase of the wave transmission coefficient (see~\eqref{eq:transmission}) is defined ambiguously as the argument of complex number. Hereafter, we study the phase $\varphi_{red}$ reduced to the interval $\left[0,2\pi\right]$. We also consider total phase of the transmitted wave, which is the phase reconstructed "by transmission", i.e., assuming that the phase of the wave increases monotonically when a small section is added to the system\footnote{This phase reconstruction method is equivalent to the "frequency" reconstruction method, when it is assumed that the phase of the transmission coefficient increases monotonically as the frequency of the incident wave increases.}.

The distribution of the total phase of the transmission coefficient for a sufficiently long disordered system is approximately Gaussian~\cite{24merzlikin2020}. The variance of the distribution increases with the length of the system L as $\sigma^2\sim L$~\cite{24merzlikin2020}. As result the corresponding distribution of the reduced phase should converge to the uniform distribution. The phase randomization approximation implies that convergence of the phase distribution.
\begin{figure}
	\includegraphics[width=1\linewidth]{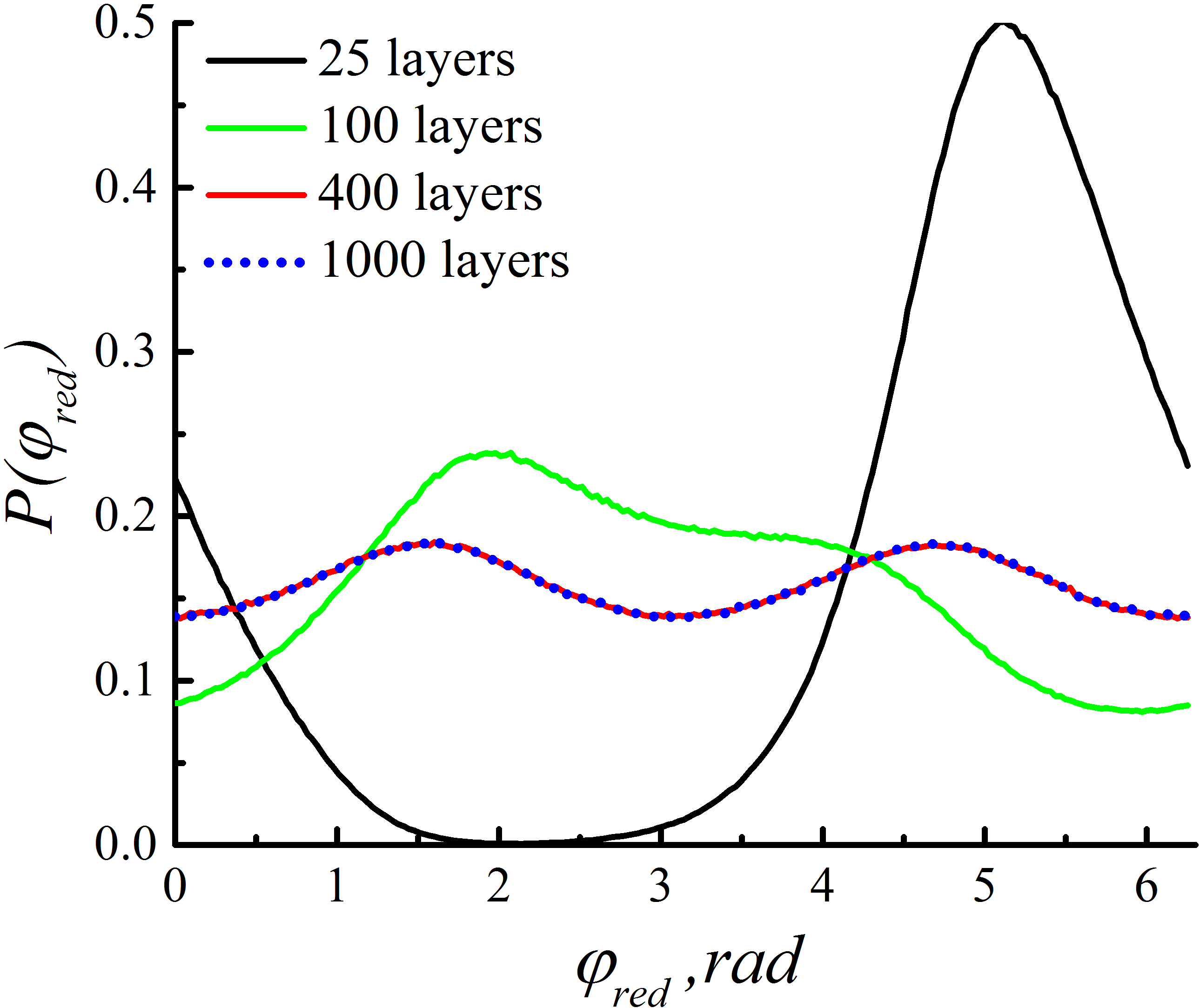}
	\caption{\label{fig:stationaryDistr} Reduced phase distribution for disordered layered systems composed of 25, 100, 400 and 1000 layers. The dielectric permittivity of the layers is uniformly distributed in the range $\left[2;4\right]$. Ensemble size is $10^7$ realizations. Thickness of each layer $k_0d=1$. Impedance of the external medium $Z_{ext}=1$.}
\end{figure}

According to numerical calculations (see Fig.~\ref{fig:stationaryDistr}) the $\varphi_{red}$ distribution converges to a certain stationary distribution with the increase of system length. However, the stationary distribution has a couple of maxima. Numerical estimates show that the magnitudes of the maxima change when the system immersed in a medium with effective impedance. However, the distribution still remains non-uniform.

\begin{figure}
	\includegraphics[width=1\linewidth]{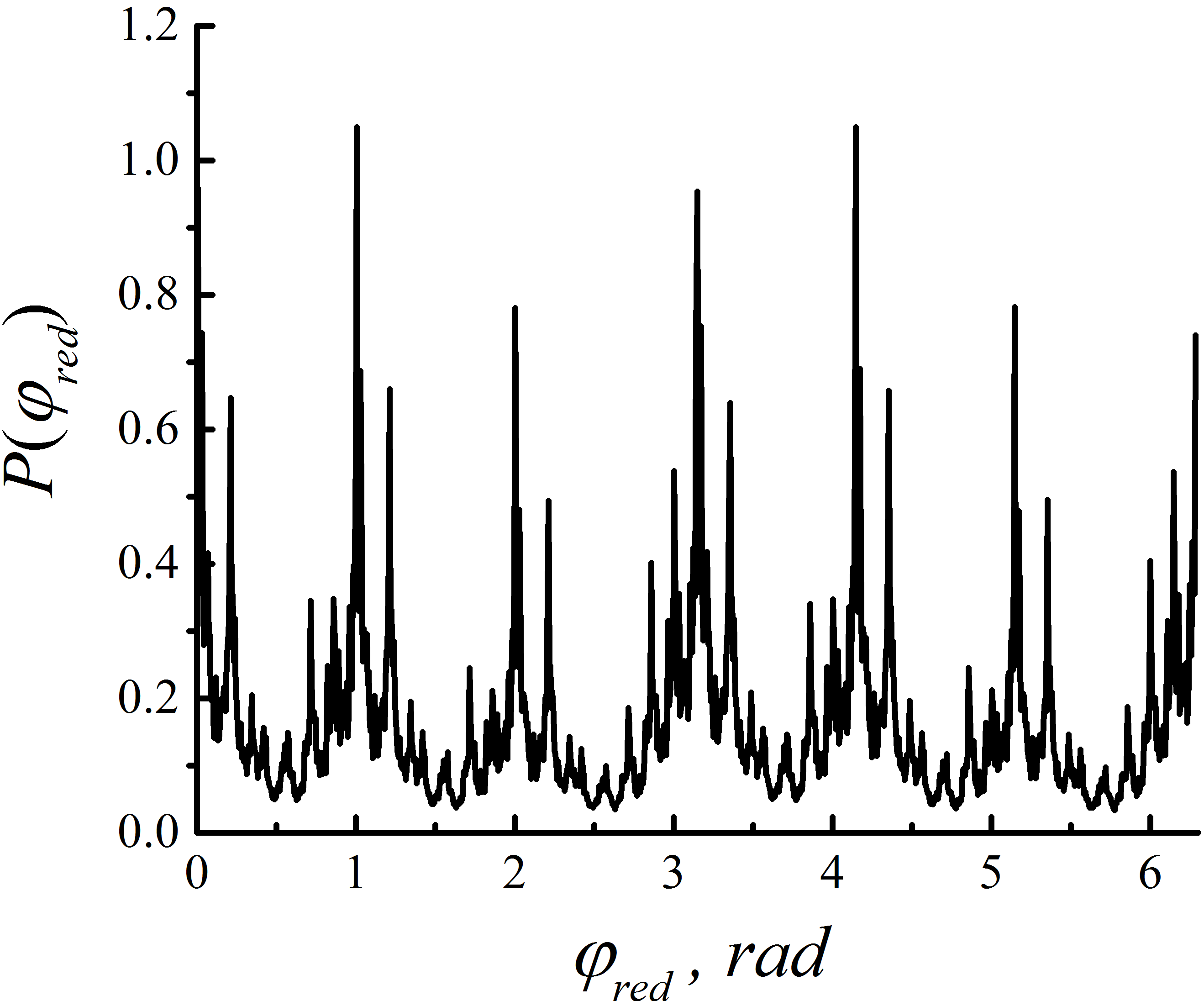}
	\caption{\label{fig:negEps} Reduced phase distribution for the system with negative contrast of dielectric permittivity. The system composed of two layer types: $\varepsilon_1=1$, $\varepsilon_2=-1$, the thickness of each layer is $k_0d=1$. The size of the system is 50 layers. Ensemble size is $10^8$ realizations.}
\end{figure}

Let us now consider a disordered layered system composed of two types of layers with dielectric permittivities $\varepsilon_1=1$ and $\varepsilon_2=-1$. The presence of layers with negative permittivity leads to a small (compared with the previous case) localization length~\eqref{eq:Lloc}: wave propagation in layers with $\varepsilon<0$ is equivalent to sub-barrier transmission in quantum mechanics. It should be noted that the system impedance cannot be matched with the external medium as soon as $\langle\varepsilon\rangle=0$. The calculations show the convergence of the phase distribution to the stationary one. In contrast to the previous case, the stationary distribution has a fractal-like shape (see Fig.~\ref{fig:negEps}) with a series of peaks, where the probability function is few times greater than for other values of $\varphi_{red}$.

The given examples show, that as a system length increases the phase distribution converge to a stationary distribution, which is non-uniform one. Thus, it is consistent to define the phase randomization as the convergence of phase distribution to stationary distribution.

\section{Phase randomization length}

Let us now consider the convergence rate of phase distribution to the stationary distribution. We will use the residual between the phase distribution $P_L\left(\varphi_{red}\right)$ for system of fixed length $L$ and the stationary distribution $P_{lim}\left(\varphi_{red}\right)$ defined according to the formula (as the measure $L_2$ between the distribution functions)
\begin{equation}
S\left(L\right)=\int_{0}^{2\pi}\left|P_L\left(\varphi_{red}\right)-P_{lim}\left(\varphi_{red}\right)\right|^2d\varphi_{red}.
\label{eq:L2measure}
\end{equation}

\begin{figure}
	\includegraphics[width=1\linewidth]{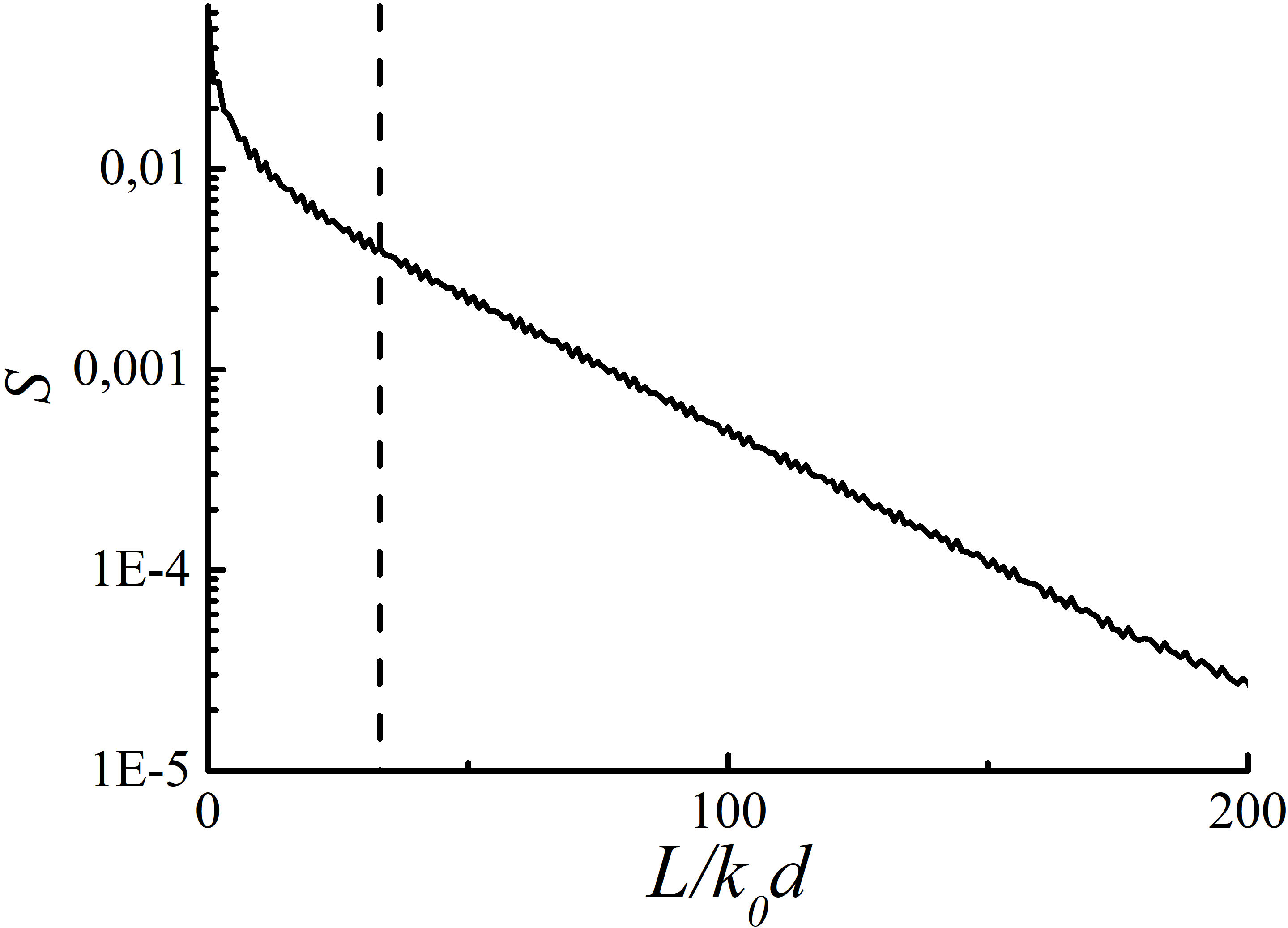}
	\caption{\label{fig:Sdecrease} Dependence of the residual~\eqref{eq:L2measure} between the distribution functions on the length of the system. System parameters correspond to Fig.~\ref{fig:stationaryDistr}. The dashed line marks the phase randomization length~\eqref{eq:exponent}.}
\end{figure}

A numerical experiment shows that $S(L)$ decreases exponentially with the number of layers (see Fig.~\ref{fig:Sdecrease})\footnote{In the case of equi-impedance disordered system (with no inner reflection) the transmission coefficient $t$ is precisely the product of transmission coefficients ${t_i}$ of the sybsystems: $t=t_N\cdot...\cdot t_0$. As a result the total phase of $t$ is the sum of $t_i$ phases. The exponential dependency of residual~\eqref{eq:L2measure} on the system length could be obtained semi-analytically in that case.}. That allows definition of the randomization length $L_{ph}$ according to
\begin{equation}
S\left(L\right)\sim \mathrm{exp}\left(-L/L_{ph}\right).
\label{eq:exponent}
\end{equation}
There should be noted that this phase randomization length is the length scale of the phase distribution convergence to the stationary distribution, which is non-uniform one.

\section{Frequency dispersion of phase randomization length}

Let us study the frequency dispersion of the phase randomization length. The dependencies~\eqref{eq:exponent} are obtained for the ensembles of systems at different frequencies. The studied layered systems are composed of the layers with the same thickness but with the dielectric permittivities chosen randomly from a uniform distribution.

\begin{figure}
	\includegraphics[width=1\linewidth]{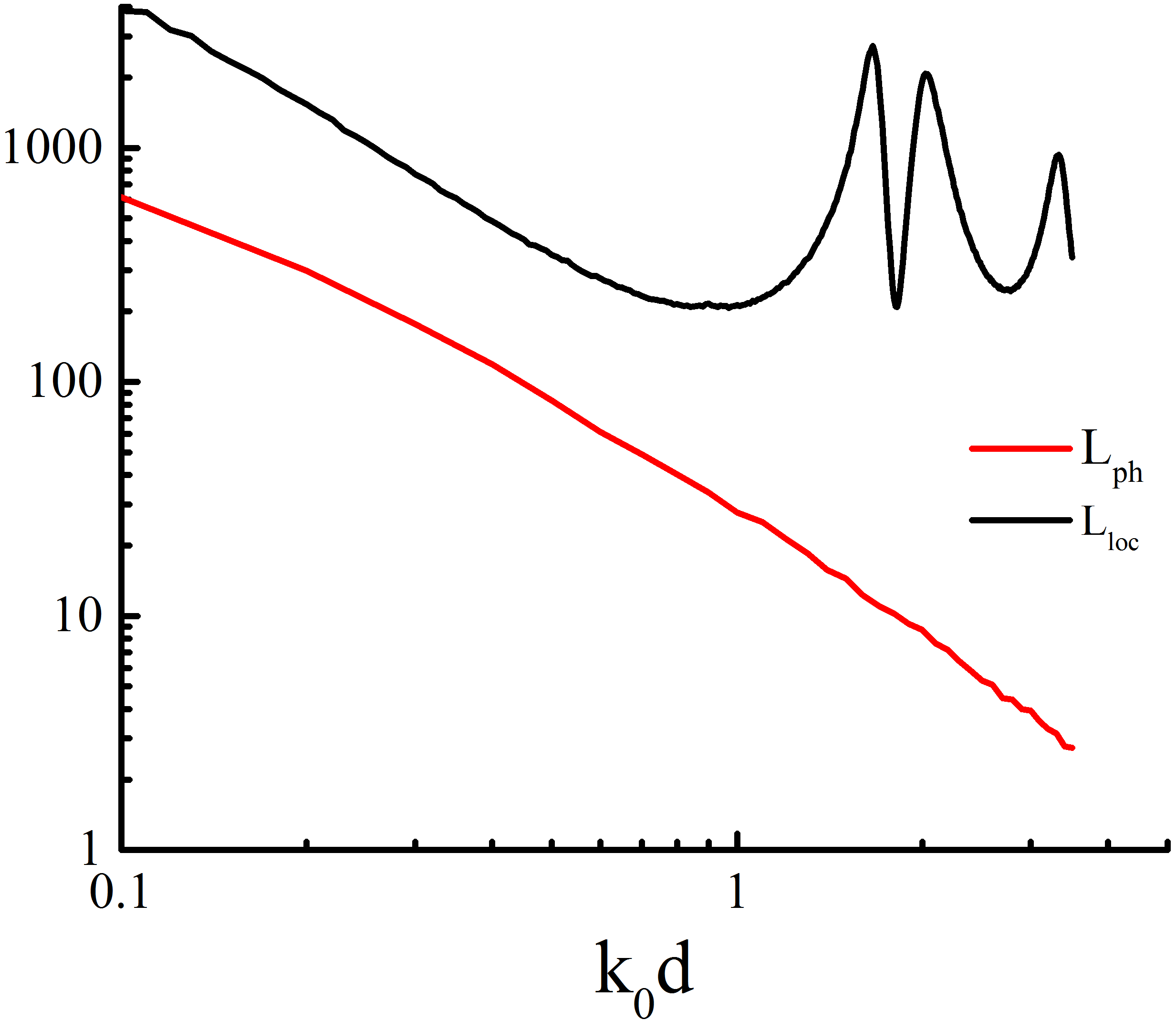}
	\caption{\label{fig:dispersion} Frequency dependences of localization length (black) and phase randomization length (red). The system parameters correspond to those given in Fig.~\ref{fig:stationaryDistr}. The size of the system is 1000 layers. Ensemble size is $10^6$ realizations. Impedance of the external medium $Z_{ext}=1$.}
\end{figure}

The frequency dispersion of phase randomization length (see Fig.~\ref{fig:dispersion}) has a power-law dependency $L_{ph}\sim\omega^\beta$, where $\beta\approx-2$. Notable, the localization length has a different frequency dependence (see Fig.~\ref{fig:dispersion}) with a number of maxima (associated with the formation of local Bragg reflectors~\cite{25vinogradov2004}) and minima. At the same time, the phase randomization length decreases monotonically with increasing frequency, which is related to a monotonic increase of the layers optical thickness and corresponding rise of the total phase distribution variance.

\section{Conclusion}

The phenomenon of phase randomization is commonly referred to in the physics of disordered systems, although the theoretical concept of the phenomenon is full of contradictions and dark spots. To clarify the theory we conducted the numerical study of the phenomenon.

Anderson localization and randomization of transmission coefficient phase can proceed independently\footnote{For example, if the layers have the same impedance, one will observe phase randomization, but not Anderson localization. This happens in a two-component mixture when falling at a Brewster angle~\cite{26sipe1988}.}, but are generally related. The phase randomization is a convergence of phase distribution to the stationary one as the system length increases. In contrast to the widely used assumption~\cite{7anderson1980} we demonstrated numerically that the stationary distribution is generally non-uniform.

The phase randomization can be characterized by the scale - the length of phase randomization $L_{ph}$, which is macroscopic quantity. The definition of the phase randomization length is given by the formula~\eqref{eq:exponent}. The frequency dispersion of the phase randomization length $L_{ph}$ is qualitatively different from the one of the localization length $L_{loc}$. The difference becomes noticeable starting with the frequencies
\begin{equation}
\langle kd\rangle\sim 1,
\label{eq:longWaves}
\end{equation}
where $d$ is the characteristic size of the inhomogeneity. The frequency dependences of $L_{ph}$ and $L_{loc}$ are similar in the region $\langle kd\rangle\ll 1$, which is possible reason for single parameter scaling applicability in that case.

\section*{Disclosure statement}

The authors declare no conflict of interest.

\section*{Funding}

The work was supported by the Foundation for the Advancement of Theoretical Physics and Mathematics "BASIS".

\end{document}